\documentclass[prd, twocolumn,  superscriptaddress, nofootinbib]{revtex4}
\usepackage{graphicx, epsfig}

\newcommand{\ApJL}{Astrophys. J. Lett.}
\newcommand{\ApJ}{Astrophys. J.}
\newcommand{\PRL}{Phys. Rev. Lett.}
\newcommand{\PRD}{Phys. Rev. D}
\newcommand{\PRP}{Phys. Rep.}
\newcommand{\MNRAS}{Mon. Not. Roy. Astron. Soc.}

\def\sun{\hbox{$\odot$}}

\long\def\comment#1{}

\def\la{\hbox{ \raise.35ex\rlap{$<$}\lower.6ex\hbox{$\sim$}\ }}
\def\ga{\hbox{ \raise.35ex\rlap{$>$}\lower.6ex\hbox{$\sim$}\ }}

\def\W2{{\cal W}}

\newcommand{\bn}{\hat{\bf n}}

\newcommand{\isw}{{\rm ISW}}

\begin{document}

\title{Growth Rate of Large Scale Structure as a Powerful Probe of Dark Energy}
\author{Asantha Cooray}
\affiliation{Theoretical Astrophysics, 
California Institute of Technology,
 Pasadena, CA 91125}
\author{Dragan Huterer}
\affiliation{Department of Physics,
Case Western Reserve University,
Cleveland, OH~~44106}
\author{Daniel Baumann}
\affiliation{DAMTP, Centre for Mathematical Sciences, University of
Cambridge, Cambridge CB3 0WA, UK}

\begin{abstract}
The redshift evolution of the growth rate of the gravitational
potential, $d(D/a)/dz$, is an excellent discriminator of dark energy
parameters and, in principle, more powerful than standard classical
tests of cosmology.  This evolution is directly observable through the
integrated Sachs-Wolfe effect in cosmic microwave background (CMB)
anisotropies. We consider the prospects of measuring the growth rate
via a novel method employed through measurements of CMB
polarization towards galaxy clusters. The potentially achievable
errors on dark energy parameters are comparable and fully
complementary to those expected from other upcoming tests of dark
energy, making this test a highly promising tool of precision cosmology.
\end{abstract}

\maketitle
%------------------------------------------------------------------------------

% User-supplied List of keywords.

%\pacs{98.80.Es,95.85.Nv,98.35.Ce,98.70.Vc
%\hfill}
%]

%\bigskip
%{\it Introduction.\hspace{0.5cm}} 
One of the key issues in modern
cosmology is developing efficient and complementary methods to measure
cosmological parameters and cosmological functions. In particular,
much interest has been devoted to developing methods to constrain the
properties of the mysterious dark energy component that causes the
recently discovered accelerated expansion of the universe~\cite{SNe}.
To this end, it has been pointed out that type Ia supernovae, number
counts, and weak gravitational lensing are all very promising probes
of the dark energy equation of state $w$ and its energy density relative
to critical $\Omega_{\rm DE}$ (see~\cite{HutTur01} and references
therein), and that a number of other methods are likely to contribute
useful information.

These cosmological tests probe various fundamental quantities. For
example, type Ia supernovae effectively measure the luminosity
distance, number counts are sensitive to a combination of the volume
element and the growth of density perturbations, while cosmic
microwave background (CMB) anisotropy effectively determines the
distance to the surface of last scattering.  While these tests are
well understood and pursued in various observational programs, the
information one can extract from these measurements is limited by the
presence of fundamental degeneracies of cosmological parameters that
enter the observable quantity in question.

It is typically advantageous when the measurements involve not the
quantity itself but rather its derivatives with respect to time or
redshift, since in that case the dependence on the equation of state
$w(z)$ is more direct. For example, the Hubble parameter $H(z)$ is
more sensitive to the equation of state than the comoving distance
$r(z)$ since $r(z)=\int dz/H(z)$, but the latter has the advantage of
being readily and accurately measurable.  In this respect, the linear
growth factor of density perturbations provides important information
since it is a function of the Hubble parameter and the equation of
state of dark energy $w$ (see below).

As we will show, the {\it rate of evolution} of the growth factor with
redshift is a tremendous tool for measuring dark energy parameters. We
will further suggest the integrated Sachs-Wolfe (ISW; \cite{SacWol67})
effect as a probe for this purpose and propose polarization
measurements of CMB anisotropy towards galaxy clusters. The latter
provides an indirect method to extract the temperature quadrupole
associated with the ISW effect as a function of the cluster redshift,
with a reduction in the cosmic variance which plagues large
scale temperature anisotropy measurements.  To make this study
practical, we consider the prospects of upcoming
arcminute scale CMB polarization observations with instruments such as
the South Pole Telescope (SPT) and the planned CMBPol satellite
mission.
%Throughout, we assume a flat universe with fraction of energy
%density in matter of $\Omega_{\rm M}=0.3$ and the rest in dark energy with
%the equation of state $w=p_{\rm DE}/\rho_{\rm DE}=-1$.

%\bigskip
%{\it Rate of Fluctuation Growth  and Dark Energy.\hspace{0.5cm}}
To begin, we review aspects related to the growth of large scale
structure.  In linear theory, all Fourier modes of the density
perturbation, $\delta (\equiv \delta \rho_{\rm M}/\rho_{rm M})$, grow at the same
rate: $\delta_k(a)=D(a) \, \delta_k(a=1)$, where $D(a)$ is the growth
factor normalized to unity today and $a=(1+z)^{-1}$ is the scale
factor. In the matter-dominated era $D(a)=a$, while in the presence of
a smooth dark energy component perturbation growth slows and $D(a)$
increases less rapidly with $a$.  In general, the growth function can
be computed by solving the linear perturbation equation
$\ddot{\delta}_k + 2 (\dot{a}/a)\dot{\delta}_k - 4 \pi G \rho_{\rm M}
\delta_k=0$ where dot is the derivative with respect to physical time.

Defining the growth suppression rate (growth rate relative to that in
a flat, matter-dominated universe) as $g(a)\equiv D(a)/a$, and still
allowing for a general $w(a)$, one can write
\begin{eqnarray}
2\frac{d^2g}{d\ln a^2} &+&\left[5 - 3w(a)\Omega_{\rm DE}(a)\right]
	\frac{dg}{d\ln a}+\nonumber \\[0.1cm] 
&+& 3\left[1-w(a)\right]\Omega_{\rm DE}(a)g=0,
\label{eqn:g}
\end{eqnarray} 
where $\Omega_{\rm DE}(a)$ is the fractional dark energy density at
the scale factor $a$.  For constant $w$, the solution is given in
terms of the hypergeometric function~\cite{Pad03}, while to compute
$g(a)$ and/or $D(a)$ for a non-constant $w(a)$ one can either solve
Eq.~(\ref{eqn:g}) numerically or use analytic approximations
\cite{WanSte98}.

It has long been known that the growth function strongly depends on
$\Omega_{\rm M}$, the fractional density in matter, and $w$. Also, the
strength of several cosmological tests, such as number
counts~\cite{Haietal00}, clustering measured in redshift slices
\cite{Cooetal01} and weak lensing
\cite{Hu_WLtomo} comes primarily from their dependence on the growth
function $D(z)$. On the other hand, it has been known that redshift or
time derivatives of distance are more directly related to dark energy
parameters; in particular, the equation of state $w(z)$ is
directly related to the first and second derivatives of distance with respect
to redshift~\cite{HutTur99}.  Unfortunately, the derivatives are not
directly measured but are obtained by taking numerical derivatives of
noisy data, which significantly increases the error in the
reconstructed $w(z)$.

\begin{figure}[t]
\includegraphics[height=3.5in, width= 2.4in, angle=-90]{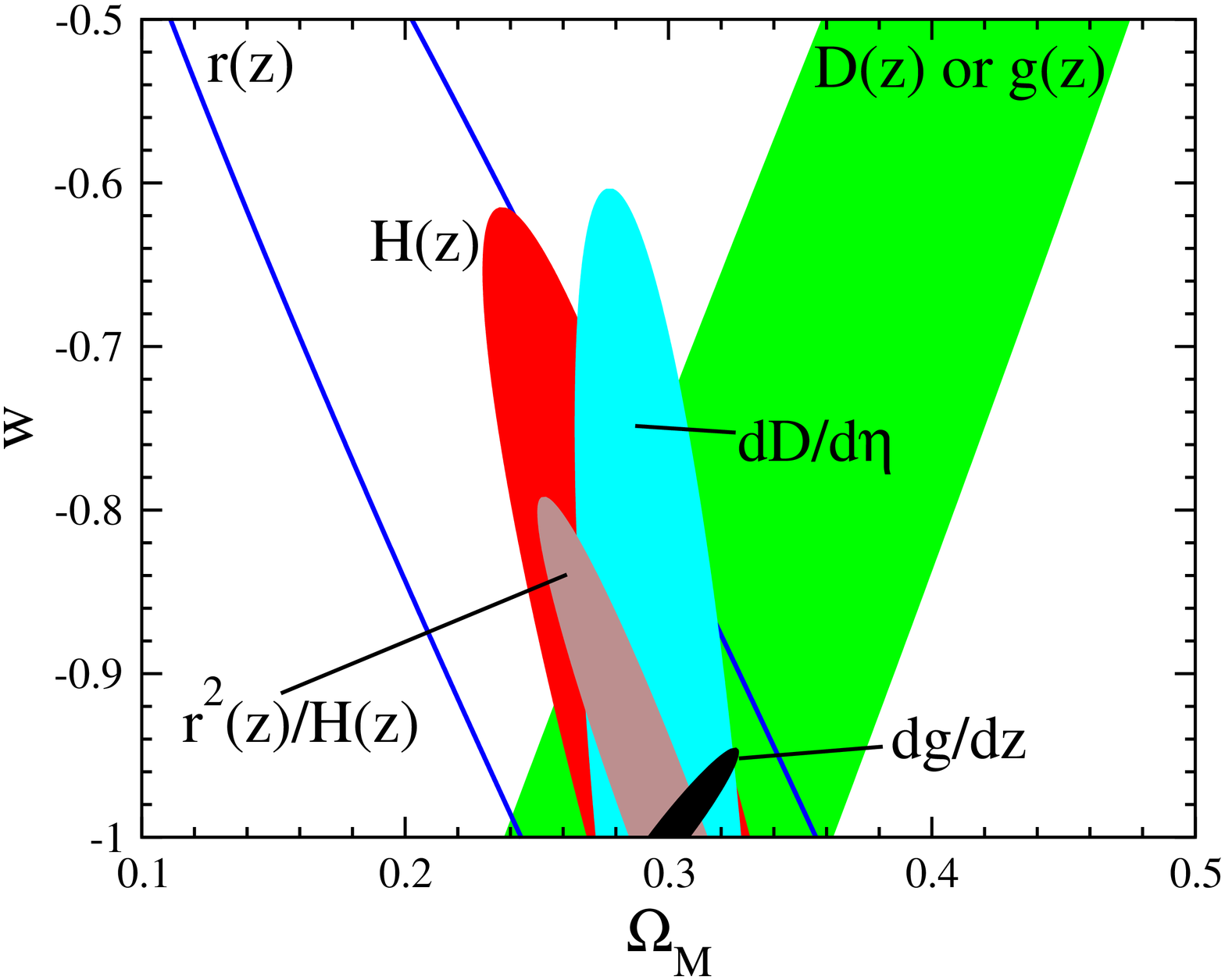}
\includegraphics[height=3.5in, width= 2.4in, angle=-90]{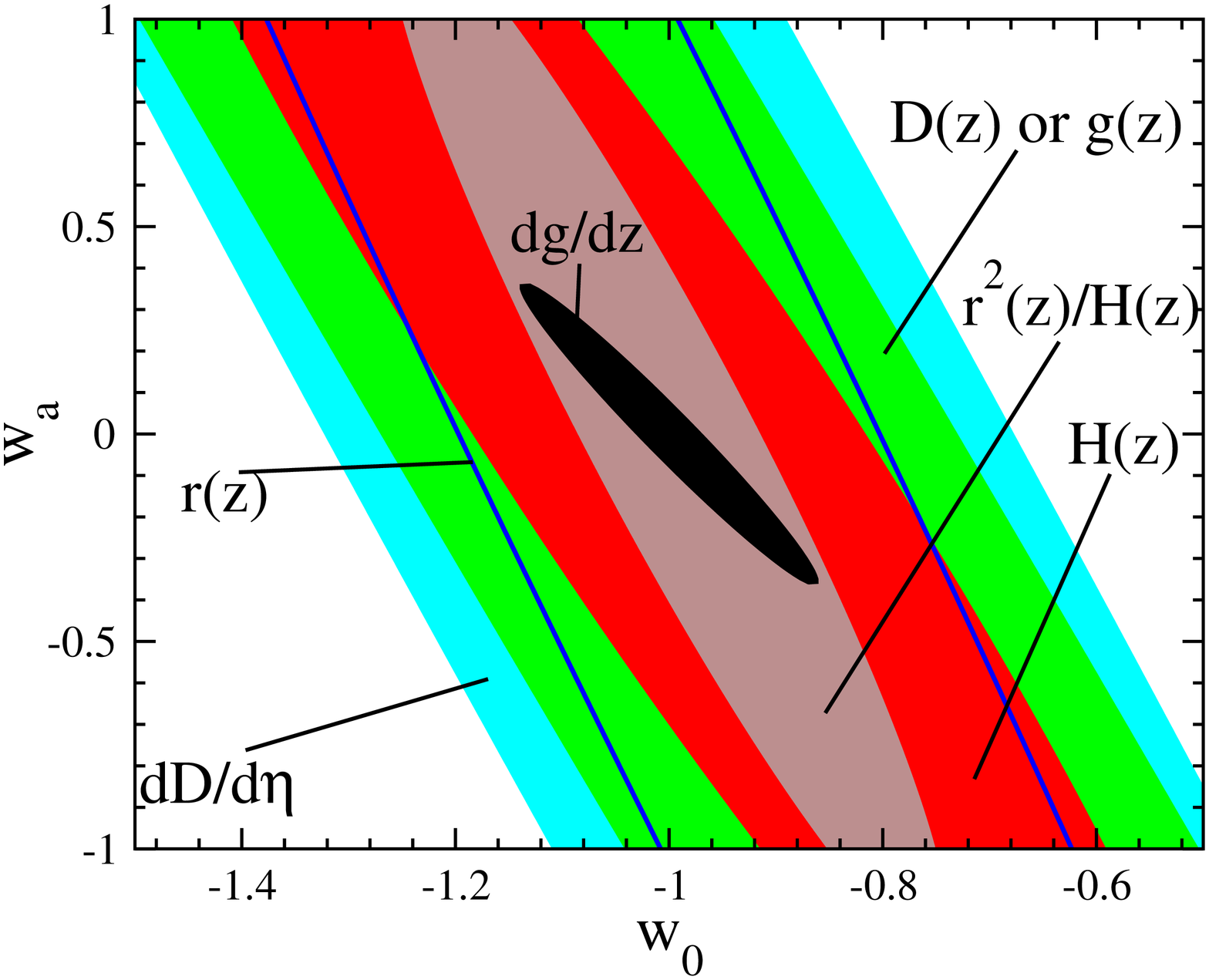}
\caption{Errors in the $\Omega_{\rm M}$--$w$ plane (top panel) and the
$w_0$--$w_a$ plane
(bottom panel, with a prior on matter density, $\sigma(\Omega_{\rm M})$ 
of 0.01) 
assuming a 10\% measurement of a given quantity at
redshifts $z=[0.1, 0.2, \ldots, 2.0]$.  
We show errors for distance $r(z)$, Hubble
parameter $H(z)$, growth factor $D(z)$, differential volume element $r^2(z)/H(z)$,
$dD/d\eta$ ($\eta$ is the conformal time), and rate of growth suppression, $dg/dz$. 
Note that the latter, $dg/dz$, is by far the most sensitive to $\Omega_{\rm M}$ and $w$. For example, a 10\% accuracy
measurement of $dg/dz$ is comparable and complementary to $<1$\%
measurement of distance.}
\label{fig:probes}
\end{figure}

It is interesting to examine the sensitivity of the {\it rate} of change of
the growth suppression factor to $\Omega_{\rm M}$ and $w$. To do this, we
first consider constant $w$, and then a two-parameter description of
time-varying $w$.  For the latter we do not choose the commonly used
$w(z)=w_0+w'z$~\cite{CooHut99}, but rather
$w(z)=w_0+w_az/(1+z)$~\cite{Lin03} which is bounded at high redshift
and facilitates the integration of Eq.~(\ref{eqn:g}). [For $w'$
aficionados, we mention that the error in $w_a$ is roughly twice the
error in $w'$.]  Fig.~\ref{fig:probes} shows the error bars in the
$\Omega_{\rm M}$--$w$ plane (top) and $w_0$--$w_a$ plane (bottom)
using various classical tests assuming a fiducial model of $\Omega_{\rm M}=0.3$ and $w=-1$; we use the same fiducial model throughout the paper.
The calculation uses the Fisher matrix formalism and assumes 10\% measurements in a given quantity at each
interval of $\Delta z=0.1$ in redshift between $z=0.1$ and $z=2$.  We
show a variety of quantities, including the distance $r(z)$, the volume
element $r^2(z)/H(z)$ and the growth factor $D(z)$, which are the most
commonly considered probes of dark energy.  We also consider $dD/d\eta$
($\eta$ is conformal time) and $dg/dz$.  As emphasized in
Ref.~\cite{PeeKno02}, $dD/d\eta$, which is measured by large-scale
velocities, is mostly sensitive to $\Omega_{\rm M}$ and not $w$.  What
Fig.~\ref{fig:probes} illustrates is that $dg/dz$ is much more
powerful than other probes due to the specific way the degeneracy is
broken. For example, for the same relative accuracy in observations,
$dg/dz$ is about 15 times stronger than the comoving distance $r(z)$! 
Of course, this comparison is not necessarily fair, since enormous
amount of work has gone into developing methods to determine
distances, which are now expected to be measurable to an accuracy of
about 1\% (per interval of 0.1 in redshift) by SNAP~\cite{SNAP},
making them the most direct probes of the cosmological expansion
history, while not much attention has been devoted to the more
esoteric quantity $dg/dz$.  In the remainder of this paper we show
that there indeed exists a very promising cosmological test which is
sensitive to $dg/dz$.

%\bigskip
%{\it Temporal Evolution of Growth through the ISW
%Effect.\hspace{0.5cm}} 
The above discussion indicates that it would be
ideal to have a cosmological probe of the evolution of growth
suppression, $dg/dz$. It turns out that just such a probe exists in a
universe that is not matter-dominated at late times.  The dark energy
domination causes the time-variation of the gravitational potential,
which in turn contributes to CMB anisotropies through the ISW effect~\cite{SacWol67}.  The resulting
temperature fluctuation is given by
\begin{equation}
\Delta T^\isw(\bn) = -2 \int_0^{r_{\rm rec}} dr'\, 
	{d\Phi (r')\over dr'} \, ,
\end{equation}
where $r_{\rm rec}$ is the radial comoving distance to last scattering
with $z_{\rm rec}=1100$.  From Poisson's equation,
$\nabla^2\Phi=3/2\,H_0^2\,\Omega_{\rm M} (\delta/a)$, it follows that
the gravitational potential $\Phi$ is proportional to the growth
suppression $g$. The ISW effect therefore gives a direct measure of
the integral of $dg/dr$ (or $dg/dz$) computed over some effective time
(or redshift) interval.

While the ISW effect determined at the present time can be used
as a probe of dark energy~\cite{Coretal03}, its contribution to
CMB temperature fluctuations is dwarfed by the primordial anisotropy
contribution at last scattering.  
Though the cross-correlation between the large scale structure
and CMB anisotropy fluctuations has been considered
as a method to extract the ISW contribution \cite{CriTur96}, 
such correlations are affected by the
dominant noise contribution related to primary anisotropies \cite{Coo02}.

%\bigskip
%{\it Dark Energy from Polarization of Clusters.\hspace{0.5cm}} 
There is another way of extracting information captured in the ISW
effect: through the measurement of CMB polarization towards galaxy
clusters \cite{CooBau03}. The polarization signal is generated by rscattering
of the temperature quadrupole seen by free electrons in the cluster frame
\cite{Challinor}. Provided that 
the optical depth to scattering in individual
clusters is determined {\it a priori} by other methods, such as the
Sunyaev-Zel'dovich (SZ; \cite{SunZel80}) effect, one can measure the
quadrupole at the cluster location with a reduction in cosmic variance 
\cite{KamLoe97}.  Note that the
quadrupole measured from a cluster at high redshift is not the same 
quadrupole as one observes today due to the difference in the
projected length scales. Since the ISW effect contributes a
significant fraction of the quadrupolar anisotropy at late times,
cluster polarization provides an indirect probe of dark energy.  Because
clusters can be selected over a wide range in
redshift, the polarization signal can be measured as a function of
redshift and inverted to reconstruct the evolution of the ISW
quadrupole \cite{CooBau03}. 

The anisotropy quadrupole, $C_2(z)$, has two contributions: one at the
surface of last scattering due to the Sachs-Wolfe (SW) effect, $C_2^{\rm
SW}(z)$, and another at late times due to the ISW effect, $C_2^{\rm
ISW}(z)$. We write these two contributions to the power spectrum,
projected to a redshift $z$, respectively as
\begin{eqnarray}
C_2^{\rm SW}(z) &=& \frac{4\pi}{9}
\int_0^{\infty} \frac{dk}{k} \Delta^2_{\Phi\Phi}(k,r_{\rm rec}) 
	j_2^2[k(r_{\rm rec}-r)] \; \; {\rm and} \nonumber\\[0.1cm]
C_2^{\rm ISW}(z) &=& 16\pi\int_0^{\infty}
	\frac{dk} {k} \Delta^2_{\Phi\Phi}(k,r_{\rm rec})\times  \\[0.1cm]
	&\times&\left[\int_{r}^{r_{rec}} dr' \frac{1}{g(z_{\rm rec})}
	\frac{d}{dr'} g(z') \ j_2(k(r'-r))\right]^2 \, . \nonumber 
\label{eqn:cl}
\end{eqnarray}
Here $r$ is the radial comoving distance out to redshift $z$ and
$\Delta^2_{\Phi\Phi}(k,r_{\rm rec}) (\equiv
k^3P_{\Phi\Phi}(k,r_{\rm rec})/2\pi^2)$ is the logarithmic power
spectrum of fluctuations in the potential field at the last scattering
surface. We will concentrate on the dark energy properties, whose
effects are dominant at low redshifts, and assume that the parameters that
define the power spectrum, such as the normalization, spectral tilt,
and physical matter and baryon densities $\Omega_{\rm M}h^2$ and
$\Omega_{\rm B}h^2$, are known to the accuracy expected from the Planck
mission with polarization information~\cite{HETW}. Given these priors, the
SW contribution is then known to a few percent accuracy. 
%For simplicity, we do not consider the so-called early quintessence
%which would modify the power spectrum at high redshift. 
Also note that, conveniently, only the large scales in the power
spectrum contribute to the ISW effect, so that we do not need to
consider thorny issues related to small-scale non-linear structures
and additional parameters such as the neutrino mass.

\begin{figure}[!ht]
\includegraphics[height=3.0in, width=1.5in, angle=-90]{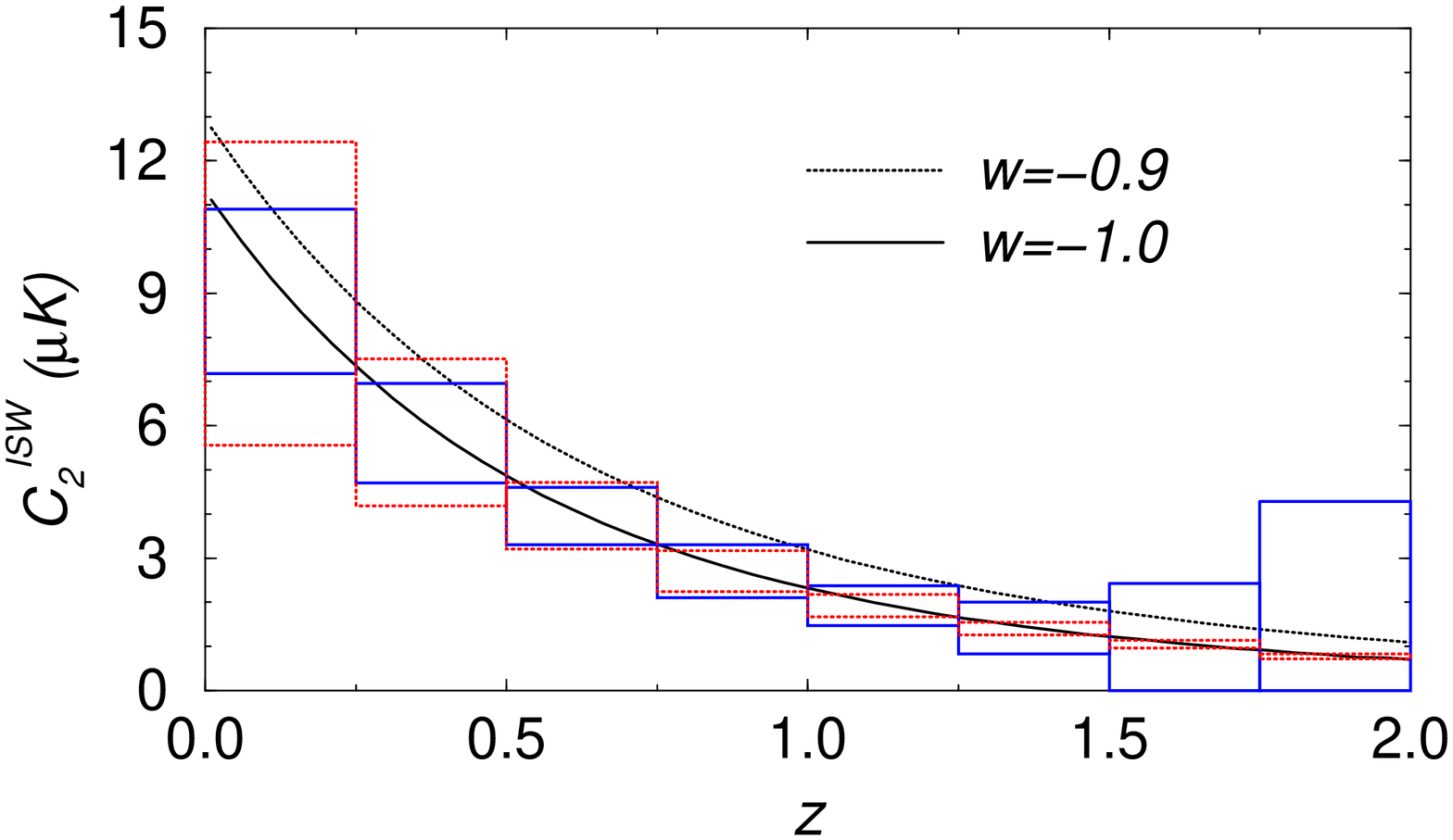}
\\[0.20cm]
\includegraphics[height=1.8in, width=1.5in, angle=-90]{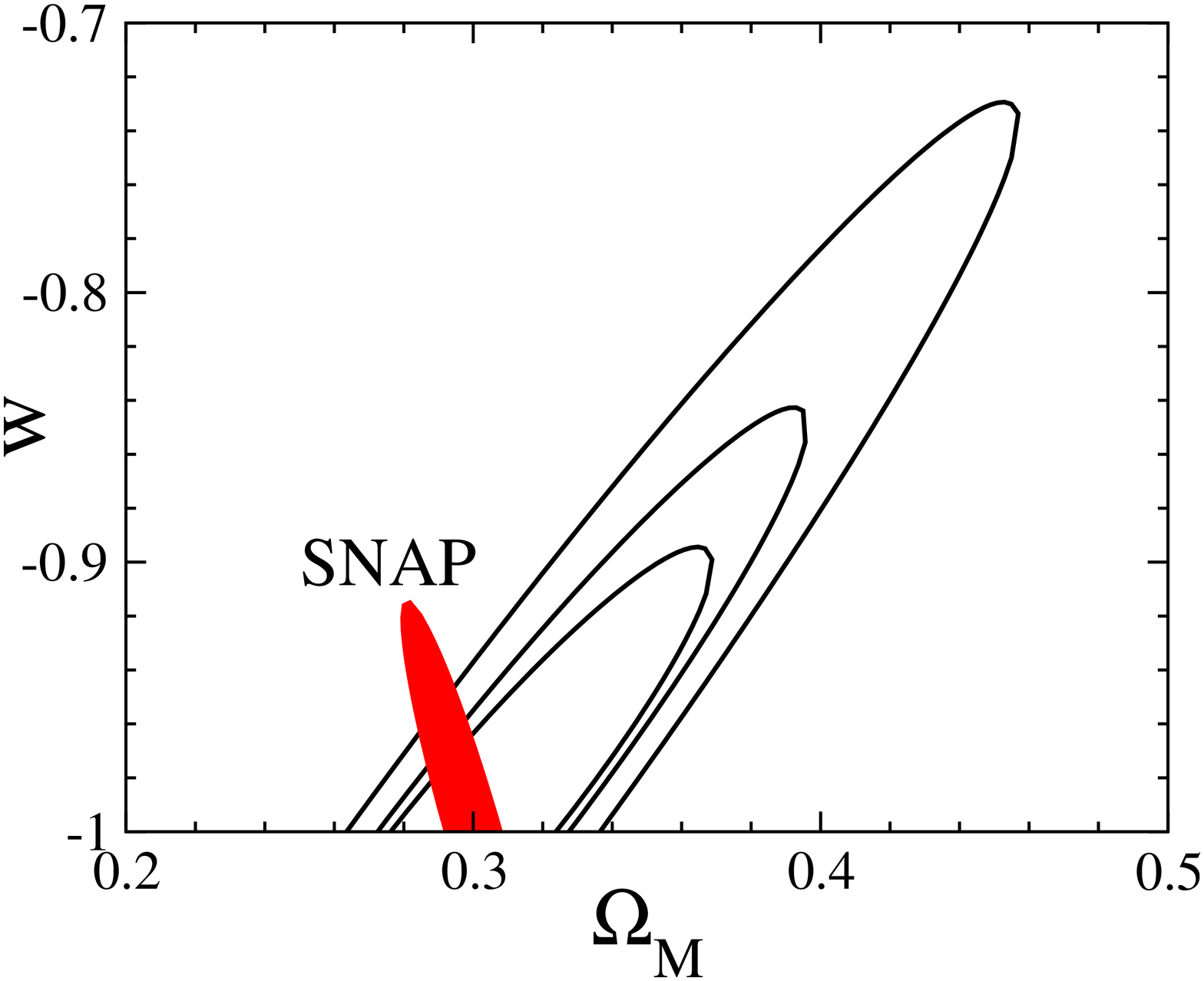}\nobreak
	\hspace{-0.4cm}
\includegraphics[height=1.8in, width=1.5in, angle=-90]{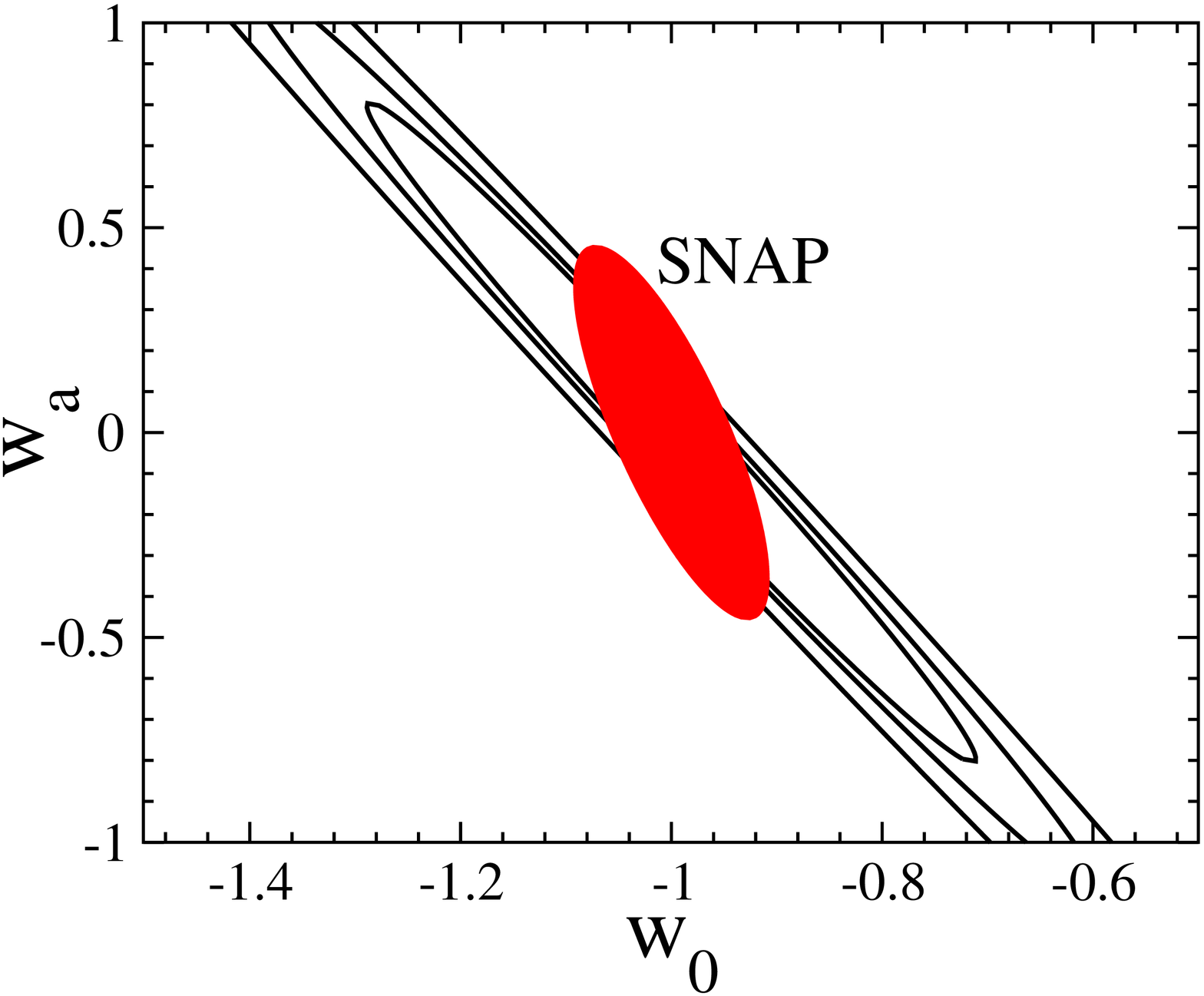}
\caption{Top panel: the projected ISW quadrupole as a function of redshift. 
The solid error bars assume a reconstruction with clusters down to
10$^{14}$ M$_{\sun}$ in an area of 10$^4$ deg$^2$ with an instrumental
noise of 0.1 $\mu$K. The dotted lines show the cosmic-variance for an
all-sky reconstruction computed from the number of independent volumes
sampled by clusters at each redshift bin~\cite{KamLoe97}.  Bottom
left: Parameter errors from the projected ISW quadrupole measurements,
assuming $w={\rm const}$. The small ellipse is for the case shown in
top panel (with cosmic variance added in quadrature), while the two
larger ellipses assume a factor of 3 and 10 increase in the
instrumental noise contribution, respectively.  For comparison, we
also show the constraints expected from SNAP. When the most optimistic 
polarization information is added, SNAP's constraints on $w$ improve by a factor of 3.
Bottom right: same, but
for $w_0$ and $w_a$ and assuming an additional prior
$\sigma(\Omega_{\rm M})=0.01$.}
\label{fig:isw}
\end{figure}

The galaxy cluster polarization signal arises from the rescattering of
the quadrupole which receives a contribution from $C_2^{\rm ISW}(z)$ at low redshifts.
Ref.~\cite{CooBau03} discussed how well this quadrupole can be
measured as a function of redshift with Planck and a ground-based
experiment with significant instrumental noise. In the top panel of
Fig.~\ref{fig:isw} we show the 
projected ISW contribution to the temperature quadrupole
as a function of redshift, and expected errors for a
ground-based survey targeting clusters down to a mass limit of
$10^{14}M_{\sun}$ in a total area of 10$^4$ deg$^2$ with an
instrumental noise for polarization observations of 0.1 $\mu$K. As in
\cite{CooBau03}, we assume four channels for these observations so
that the ISW quadrupole can be separated from the contribution of the
kinematic quadrupole. The latter has a distinct spectrum and the
separation based on frequency information leads to an overall increase
in noise by a factor of 2 to 3 depending on the exact frequencies of
channels selected.  Note that we have assumed an instrumental noise of
0.1 $\mu$K for these observations.  While a polarization sensitive
detector array on the SPT can be expected to reach noise
levels of $\sim$ 1 $\mu$K or less per pixel, we have assumed an order
of magnitude reduction in noise, as expected from the
planned CMBPol satellite mission.
Since the expected noise level for arcminute scale
polarization observations from such a mission is not currently
defined, and to consider ground-based efforts such as the SPT, we have
considered the range of values between 0.1 and 1 $\mu$K so as to
obtain some guidance on how well cluster polarization measurements
with noise in this range can be used to probe dark energy.

In addition to instrumental noise, the polarization measurements are
subject to cosmic variance. This variance is determined by the number of
independent volumes that last scattering spheres of individual 
clusters, in some redshift bin, occupy 
\cite{KamLoe97}. Dotted lines in the top panel of
Fig.~\ref{fig:isw} show the cosmic variance contribution for an all-sky
experiment. As one moves to higher redshift, the number of independent
samplings of the local quadrupole increases, leading to a reduction in
cosmic variance.  The expected redshift distribution
of clusters peaks at redshifts around 1--1.5 where it provides the
best estimate of the local quadrupole, while errors increase at very
low and high redshift due to the smaller number of clusters.

To consider how well these observations can be used to understand dark
energy parameters, we again perform a Fisher matrix calculation.  The
bottom panels of Fig.~\ref{fig:isw} show how well $\Omega_{\rm M}$ and
$w$ (assuming a flat universe and constant $w$), and $w_0$ and $w_a$
(assuming a two-parameter description of $w(z)$ as before and a prior
on $\Omega_{\rm M}$ of 0.01) can be measured.  While the errors are
fairly large with a 1 $\mu$K noise level per pixel, improving this
noise threshold to 0.1 $\mu$K leads to significant gains in the determination of
$\Omega_{\rm M}$ and $w$. Note also that these errors roughly scale as the
inverse square root of the area of sky covered, and with all-sky
coverage the errors are expected to decrease by a factor of two. With
an order of magnitude improvement in noise, the redshift evolution of
the ISW effect extracted from polarization measurements becomes a
powerful probe of dark energy providing significant estimates of
parameters, comparable and complementary to type Ia supernovae.

%\bigskip
%{\it Conclusions.\hspace{0.5cm}} 
To conclude, we have argued that the rate of evolution of the growth
suppression factor, $dg/dz$, is a very powerful probe of dark energy.
We have shown that the polarization signal from a large number of galaxy
clusters is directly related to this quantity, and can be used to constrain
dark energy parameters. In the next decade, the
planned mission CMBPol is expected to reach a sensitivity of order 0.1
$\mu$K at arcminute resolution and have all-sky coverage, providing
polarization measurements of a significant number $(\sim 10^4)$ of clusters, 
 from which the quadrupole can be reconstructed as a function of redshift.
Although our study is preliminary, we have shown that this method can
provide constraints on the dark energy equation of state and its time
variation comparable and complementary to those from type Ia
supernovae and other well-studied probes of dark energy.  More
importantly, this method is entirely different from most of the others
both in its theoretical underpinnings and in the systematic errors
expected.  Combining this method with others opens the exciting
possibility of significantly improving the constraints on $w$ and
helps usher a new era in our exploration of dark energy.

We thank I. Maor, A. Melchiorri and J. Ruhl for useful discussions.
This work was supported in part by DoE (AC and DH) and the Sherman
Fairchild foundation (AC). AC thanks the Particle Astrophysics Group
at CWRU for hospitality while this work was initiated.

\end{document}